\documentclass[twocolumn,showpacs,amsmath,amssymb,prl,superscriptaddress,nofootinbib]{revtex4}

\usepackage{graphicx}

\begin{document}
\title{Bound entanglement provides convertibility of pure entangled states}
\author{Satoshi Ishizaka}
\affiliation{
PRESTO, Japan Science and Technology Agency,
4-1-8 Honcho Kawaguchi, Saitama, Japan}
\affiliation{
Fundamental Research Laboratories, NEC Corporation, 
34 Miyukigaoka, Tsukuba, Ibaraki, Japan}
\date{\today}
%%%%%%%%%%%%%%%%%%%%%%%%%%%%%%%%%%%%%%%%%%%%%%%%%%%%%%%%%%%%%%%%%%%%%%%%%%%%%%%
\begin{abstract}
I show that two distant parties can transform pure
entangled states to arbitrary pure states by stochastic local
operations and classical communication (SLOCC) at the single copy level,
if they share bound entangled states.
This is the effect of bound entanglement since this entanglement
processing is impossible by SLOCC alone.
Similar effect of bound entanglement occurs in three qubits
where two incomparable entangled states of GHZ and W 
can be inter-converted.
In general multipartite settings composed by $N$ distant parties,
all $N$-partite pure entangled states are inter-convertible
by SLOCC with the assistance of bound entangled states with positive partial
transpose.
\end{abstract}
\pacs{03.67.Mn, 03.65.Ud}
%%%%%%%%%%%%%%%%%%%%%%%%%%%%%%%%%%%%%%%%%%%%%%%%%%%%%%%%%%%%%%%%%%%%%%%%%%%%%%%
\maketitle
\begin{figure}
\centerline{\scalebox{0.35}[0.35]{\includegraphics{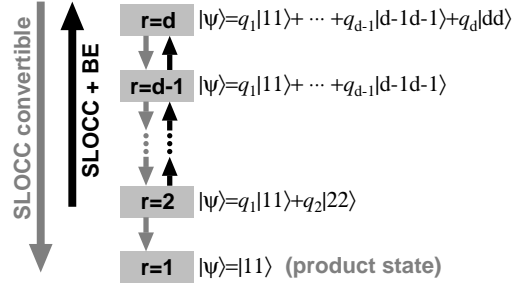}}}
\caption{
The transformation of a single copy of pure states by SLOCC is possible
only in the decreasing direction of the Schmidt rank ($r$). 
This restriction is largely removed
by the assistance of bound entangled (BE) states.
}
\label{fig: 1}
\end{figure}
\begin{figure}[t]
\centerline{\scalebox{0.4}[0.4]{\includegraphics{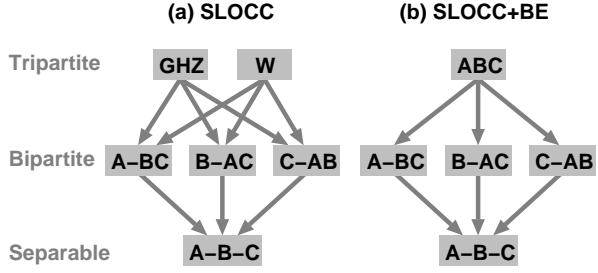}}}
\caption{
The classification and convertibility of pure states in three qubits
by (a) SLOCC \cite{Dur00b} and
(b) SLOCC with the assistance of PPT-BE states.
Two incomparable classes of tripartite entanglement (GHZ and W) are merged
into a single class (ABC) by the assistance of PPT-BE states.
}
\label{fig: 2}
\end{figure}

%%%%%%%%%%%%%%%%%%%%%%%%%%%%%%%%%%%%%%%%%%%%%%%%%%%%%%%%%%%%%%%%%%%%%%%%%%%%%%%
In quantum entanglement processing, the transformation of
entangled states by local operations and classical communication (LOCC)
is a basic task, where many intriguing aspects concerning convertibility
and irreversibility appear.
All bipartite pure entangled states are inter-convertible in the asymptotic
transformation \cite{Bennett96b} where infinitely many identical copies
of states are processed. 
Therefore, all bipartite pure entangled states can be used to perform
the same task of entanglement processing in the asymptotic regime. 
However, there exists a restriction in the
transformation of a single copy of bipartite pure states
(Fig.\ \ref{fig: 1}):
two distant parties cannot increase the number of
superposed terms (Schmidt rank, the rank of the reduced density matrix)
by LOCC even in a stochastic manner (such stochastic LOCC is denoted by SLOCC)
\cite{Lo97a,Nielsen99a,Vidal99a}.
As a result, bipartite entanglement is classified by the Schmidt rank
from the viewpoint of the convertibility at the single copy level
\cite{Dur00b}.
Such restriction becomes more strict in multipartite
settings.
In three qubits, there are two different types of tripartite entanglement:
GHZ and W type \cite{Dur00b}
[the GHZ state is $(|111\rangle\!+\!|222\rangle)/\sqrt{2}$
and W state is
$(|112\rangle\!+\!|121\rangle\!+\!|211\rangle)/\sqrt{3}$].
These cannot be transformed to each other by SLOCC,
and such entangled states are said to be incomparable (Fig.\ \ref{fig: 2}).
In general multipartite settings composed by $N$ distant parties, there are
many (possibly infinitely many) incomparable types of $N$-partite entanglement.
\par
%%%%%%%%%%%%%%%%%%%%%%%%%%%%%%%%%%%%%%%%%%%%%%%%%%%%%%%%%%%%%%%%%%%%%%%%%%%%%%%
%%
%%%%%%%%%%%%%%%%%%%%%%%%%%%%%%%%%%%%%%%%%%%%%%%%%%%%%%%%%%%%%%%%%%%%%%%%%%%%%%%
On the other hand, a remarkable aspect of the irreversibility is the existence
of bound entangled (BE) states \cite{Horodecki98a}.
Distant parties need to consume pure entangled states to prepare BE states,
but they cannot distill pure entangled states from it any more.
Much attention has been paid to this weak type of entanglement to
clarify its properties and usefulness for quantum information processing.
BE states by itself are useless for both quantum teleportation
\cite{Horodecki99a} and superdense coding \cite{Horodecki01c}.
However, it has been shown that BE states can activate the bound entanglement
of the other state \cite{Horodecki99c} and are useful for secure key
distribution \cite{Horodecki03a} in bipartite settings.
In multipartite settings, remote information concentration \cite{Murao01a},
violation of Bell's inequality \cite{Dur01a},
superactivation \cite{Shor03a}, and superadditivity of quantum capacity
\cite{Dur04a} have been reported.
However, most of these effects of BE states are concerning the entanglement
processing for mixed states.
\par
%%%%%%%%%%%%%%%%%%%%%%%%%%%%%%%%%%%%%%%%%%%%%%%%%%%%%%%%%%%%%%%%%%%%%%%%%%%%%%%
%%
%%%%%%%%%%%%%%%%%%%%%%%%%%%%%%%%%%%%%%%%%%%%%%%%%%%%%%%%%%%%%%%%%%%%%%%%%%%%%%%
%%%%%%%%%%%%%%%%%%%%%%%%%%%%%%%%%%%%%%%%%%%%%%%%%%%%%%%%%%%%%%%%%%%%%%%%%%%%%%%
%%%%%%%%%%%%%%%%%%%%%%%%%%%%%%%%%%%%%%%%%%%%%%%%%%%%%%%%%%%%%%%%%%%%%%%%%%%%%%%
In this paper, I show that BE states strongly influence the entanglement
processing of pure states at the single copy level. 
Two distant parties can get capability to
increase the Schmidt rank of bipartite pure entangled states
(in a stochastic manner) as large as they desire 
by the assistance of BE states (Fig.\ \ref{fig: 1}).
Similar effect of bound entanglement occurs in three qubits
where two incomparable entangled states of GHZ and W 
can be inter-converted (Fig.\ \ref{fig: 2}).
In general, all $N$-partite pure entangled states are
inter-convertible, and hence there is only one type of $N$-partite
entanglement from the viewpoint of SLOCC assisted by BE states.
\par
%%%%%%%%%%%%%%%%%%%%%%%%%%%%%%%%%%%%%%%%%%%%%%%%%%%%%%%%%%%%%%%%%%%%%%%%%%%%%%%
%%
%%%%%%%%%%%%%%%%%%%%%%%%%%%%%%%%%%%%%%%%%%%%%%%%%%%%%%%%%%%%%%%%%%%%%%%%%%%%%%%
%%%%%%%%%%%%%%%%%%%%%%%%%%%%%%%%%%%%%%%%%%%%%%%%%%%%%%%%%%%%%%%%%%%%%%%%%%%%%%%
%%%%%%%%%%%%%%%%%%%%%%%%%%%%%%%%%%%%%%%%%%%%%%%%%%%%%%%%%%%%%%%%%%%%%%%%%%%%%%%
Let us first consider bipartite settings.
A state is called positive partial transpose (PPT) state
if the partially transposed density matrix remains positive, i.e.
$\varrho^{T_A}\!\ge\!0$ \cite{Peres96a}.
PPT states are undistillable, and hence
entangled PPT states are BE states \cite{Horodecki98a}.
The effects of such PPT-BE states can be taken into account by
considering the PPT maps which preserve the positivity of the partial
transpose 
\cite{Rains99b,Rains01a,Eggeling01a,Vollbrecht02a,Audenaert03a,Cirac01a}.
Among such PPT maps, stochastic PPT maps of
non-trace-preserving are considered.
Suppose that two distant parties (A and B) wish to accomplish the stochastic
transformation described by the non-trace-preserving map of
$\varrho\!\rightarrow\!S(\varrho)$.
Let $\Gamma$ be the map of the partial transpose with respect to the party A as
$\Gamma(X)\!=\!X^{T_A}$.
In this paper, $S$ is called a stochastic PPT map (SPPT map)
when both $S$ and $\Gamma\!\circ\!S\!\circ\!\Gamma$ are
completely positive (CP) maps \cite{Rains99b}.
The SPPT maps defined in this way can be always
implemented by SLOCC with the assistance of PPT-BE states
as explicitly shown later.
\par
%%%%%%%%%%%%%%%%%%%%%%%%%%%%%%%%%%%%%%%%%%%%%%%%%%%%%%%%%%%%%%%%%%%%%%%%%%%%%%%
%%
%%%%%%%%%%%%%%%%%%%%%%%%%%%%%%%%%%%%%%%%%%%%%%%%%%%%%%%%%%%%%%%%%%%%%%%%%%%%%%%
Let $S$ be an SPPT map which transforms 
$\varrho$ on ${\mathbb C}^{m}\otimes{\mathbb C}^{m}$
to a maximally entangled state
$P^+_d\!\equiv\!|\phi^+_d\rangle\langle\phi^+_d|$
on ${\mathbb C}^{d}\otimes{\mathbb C}^{d}$
where $|\phi^+_d\rangle\!=\!\sum_{i\!=\!1}^d |ii\rangle/\sqrt{d}$.
Let $T$ be a $U\otimes U^*$-twirling map of
$T(X)\!=\!\int dU (U\otimes U^*) X (U\otimes U^*)^\dagger$.
%%%%%%%%%%%%%%%%%%%%%%%%%%%%%%%%%%%%%%%%%%%%%%%%%%%%%%%%%%%%%%%%%%%%%%%%%%%%%%%
Following an idea of \cite{Rains01a,Eggeling01a},
if an SPPT map $S$ accomplishes $\varrho\!\rightarrow\!P^+_d$ 
for a given $\varrho$, the composed map
of $T\!\circ\!S$ is also an SPPT map which accomplishes 
$\varrho\!\rightarrow\!P^+_d$ with the same probability as $S$,
since $T(P^+_d)\!=\!P^+_d$ and $T$ is trace-preserving.
Therefore, one may assume $S\!=\!T\!\circ\!S$.
Since $S$ is a linear map, it must be written as
%%%%%%%%%%%%%%%%%%%%%%%%%%%%%%%%%%%%%%%%%%%%%%%%%%%%%%%%%%%%%%%%%%%%%%%%%%%%%%%
\begin{equation}
S(X)=(\hbox{tr}XA)P^+_d+(\hbox{tr}XB)\frac{I_d-P^+_d}{d^2-1},
\label{eq: SPPT}
\end{equation}
%%%%%%%%%%%%%%%%%%%%%%%%%%%%%%%%%%%%%%%%%%%%%%%%%%%%%%%%%%%%%%%%%%%%%%%%%%%%%%%
where $I_d$ is an identity operator on 
${\mathbb C}^{d}\otimes{\mathbb C}^{d}$ and
the probability of this transformation is 
%%%%%%%%%%%%%%%%%%%%%%%%%%%%%%%%%%%%%%%%%%%%%%%%%%%%%%%%%%%%%%%%%%%%%%%%%%%%%%%
\begin{equation}
P[X\!\rightarrow\!S(X)]=\hbox{tr}S(X)=\hbox{tr}X(A+B).
\end{equation}
%%%%%%%%%%%%%%%%%%%%%%%%%%%%%%%%%%%%%%%%%%%%%%%%%%%%%%%%%%%%%%%%%%%%%%%%%%%%%%%
The matrices $A$ and $B$ are chosen so that $S$ is an SPPT map.
The conditions for $A$ and $B$ are as follows:
\par
%%%%%%%%%%%%%%%%%%%%%%%%%%%%%%%%%%%%%%%%%%%%%%%%%%%%%%%%%%%%%%%%%%%%%%%%%%%%%%%
%%
%%%%%%%%%%%%%%%%%%%%%%%%%%%%%%%%%%%%%%%%%%%%%%%%%%%%%%%%%%%%%%%%%%%%%%%%%%%%%%%
{Lemma 1:}
{\it
$S$ in the form of Eq.\ (\ref{eq: SPPT}) is an SPPT map if and only if
the matrices $A$ and $B$ satisfy $A,B\!\ge\!0$, $I\!\ge\!A\!+\!B$, and
$\frac{1}{d-1}B^{T_A}\!\ge\!A^{T_A}\!\ge\!-\frac{1}{d+1}B^{T_A}$.
}
\par
%%%%%%%%%%%%%%%%%%%%%%%%%%%%%%%%%%%%%%%%%%%%%%%%%%%%%%%%%%%%%%%%%%%%%%%%%%%%%%%
%%
%%%%%%%%%%%%%%%%%%%%%%%%%%%%%%%%%%%%%%%%%%%%%%%%%%%%%%%%%%%%%%%%%%%%%%%%%%%%%%%
{Proof:} One may prove this in almost the same manner as 
\cite{Rains01a,Eggeling01a} where trace-preserving PPT maps
have been considered.
$S$ is a CP map if and only if
%%%%%%%%%%%%%%%%%%%%%%%%%%%%%%%%%%%%%%%%%%%%%%%%%%%%%%%%%%%%%%%%%%%%%%%%%%%%%%%
\begin{equation}
E_{AB}\equiv
(S_{A_1B_1} \otimes I_{A_2B_2})(P^+_{mA_1A_2} \otimes P^+_{mB_1B_2})
\ge 0,
\label{eq: state E}
\end{equation}
%%%%%%%%%%%%%%%%%%%%%%%%%%%%%%%%%%%%%%%%%%%%%%%%%%%%%%%%%%%%%%%%%%%%%%%%%%%%%%%
which leads to $A\!\ge\!0$, $B\!\ge\!0$.
Similarly, $\Gamma\!\circ\!S\!\circ\!\Gamma$ is a CP map if and only if
%%%%%%%%%%%%%%%%%%%%%%%%%%%%%%%%%%%%%%%%%%%%%%%%%%%%%%%%%%%%%%%%%%%%%%%%%%%%%%%
\begin{equation}
\big[(S_{A_1B_1} \otimes I_{A_2B_2})(P^+_{mA_1A_2} \otimes P^+_{mB_1B_2})^{T_A}\big]^{T_A}\ge0,
\label{eq: PPT condition}
\end{equation}
%%%%%%%%%%%%%%%%%%%%%%%%%%%%%%%%%%%%%%%%%%%%%%%%%%%%%%%%%%%%%%%%%%%%%%%%%%%%%%%
which leads to
$\frac{1}{d-1}B^{T_A}\!\ge\!A^{T_A}\!\ge\!-\frac{1}{d+1}B^{T_A}$.
Further, the trace condition where $P\!=\!\hbox{tr}X(A+B)\!\le\!1$ for any
input state $X$ leads to $A\!+\!B\!\le\!I$.
\hfill $\Box$
\par
%%%%%%%%%%%%%%%%%%%%%%%%%%%%%%%%%%%%%%%%%%%%%%%%%%%%%%%%%%%%%%%%%%%%%%%%%%%%%%%
%
%%%%%%%%%%%%%%%%%%%%%%%%%%%%%%%%%%%%%%%%%%%%%%%%%%%%%%%%%%%%%%%%%%%%%%%%%%%%%%%
It should be noted that, since $S$ is a CP map,
it is necessarily written in the operator-sum 
representation as $S(X)\!=\!\sum_j F_j X F_j^\dagger$, where
$F_j$'s are operation elements and satisfy $\sum_j F_j^\dagger F_j\!\le\!I$.
If only a single operation element constitutes the SPPT
map like $S(X)\!=\! F X F^\dagger$, it can be shown using
Eq.\ (\ref{eq: PPT condition}) that
$F$ must be written in a separable form as $F\!=\!G_A\otimes H_B$,
and the map turns out to be an SLOCC map.
Therefore, SPPT maps which cannot be accomplished by SLOCC
must be constituted by at least two operation elements.
\par
%%%%%%%%%%%%%%%%%%%%%%%%%%%%%%%%%%%%%%%%%%%%%%%%%%%%%%%%%%%%%%%%%%%%%%%%%%%%%%%
%%
%%%%%%%%%%%%%%%%%%%%%%%%%%%%%%%%%%%%%%%%%%%%%%%%%%%%%%%%%%%%%%%%%%%%%%%%%%%%%%%
Further, since the SPPT map I am considering here must output $P^+_d$
for a given input $\varrho$, the second term in Eq.\ (\ref{eq: SPPT})
must vanish
when $X\!=\!\varrho$.
As a result, $\hbox{tr}\varrho B\!=\!0$ must hold.
Further, $P(\varrho\!\rightarrow\!P^+_d)=\hbox{tr}\varrho A\!>\!0$
in order that the transformation is accomplished with nonzero probability.
As mentioned above, if
some SPPT map accomplishes the transformation of $\varrho\!\rightarrow\!P^+_d$,
at least one SPPT map having the form of Eq.\ (\ref{eq: SPPT}) must exist.
These two SPPT maps give the same probability,
and I obtain:
\par
%%%%%%%%%%%%%%%%%%%%%%%%%%%%%%%%%%%%%%%%%%%%%%%%%%%%%%%%%%%%%%%%%%%%%%%%%%%%%%%
%%
%%%%%%%%%%%%%%%%%%%%%%%%%%%%%%%%%%%%%%%%%%%%%%%%%%%%%%%%%%%%%%%%%%%%%%%%%%%%%%%
{Lemma 2:}
{\it
For a given $\varrho$, the stochastic transformation of
$\varrho\!\rightarrow\!P^+_d$ via SPPT maps is possible if and only
if there exist matrices $A$ and $B$ such that
$\hbox{tr}\varrho B\!=\!0$, $\hbox{tr}\varrho A\!>\!0$,
and satisfy all conditions in Lemma 1.
}
%%%%%%%%%%%%%%%%%%%%%%%%%%%%%%%%%%%%%%%%%%%%%%%%%%%%%%%%%%%%%%%%%%%%%%%%%%%%%%%
\par
%%%%%%%%%%%%%%%%%%%%%%%%%%%%%%%%%%%%%%%%%%%%%%%%%%%%%%%%%%%%%%%%%%%%%%%%%%%%%%%
%%
%%%%%%%%%%%%%%%%%%%%%%%%%%%%%%%%%%%%%%%%%%%%%%%%%%%%%%%%%%%%%%%%%%%%%%%%%%%%%%%
Then, the problem investigating the convertibility of $\varrho$ to $P^+_d$
via SPPT maps was reduced to the problem searching for the matrices
$A$ and $B$.
Let us consider the case where $\varrho$ is invariant under the
$U\otimes U^*$-twirling, and
suppose that $A$ and $B$ satisfy all conditions in Lemma 2.
Clearly, $T(A),T(B)\!\ge\!0$ and $I\!\ge\!T(A)+T(B)$.
Since $[T(X)]^{T_A}\!=\!V(X^{T_A})$
where $V$ is a $U\otimes U$-twirling map,
$\frac{1}{d-1}[T(B)]^{T_A}\!\ge\![T(A)]^{T_A}\!\ge\!-\frac{1}{d+1}[T(B)]^{T_A}$
also holds.
Further, $\hbox{tr}\varrho T(B)\!=\!\hbox{tr}T(\varrho) B\!=\!
\hbox{tr} \varrho B\!=\!0$ and $\hbox{tr}\varrho T(A)\!=\!\hbox{tr}\varrho A
\!=\!P(\varrho\!\rightarrow\!P^+_d)$.
As a result, $T(A)$ and $T(B)$ satisfy all conditions
in Lemma 2 as well as $A$ and $B$, giving the same probability.
Therefore, it suffices to consider $A$ and $B$ such that
$A\!=\!T(A)$ and $B\!=\!T(B)$, i.e. 
$A\!=\!\alpha P^+_m\!+\!\beta (I_m\!-\!P^+_m)$
and
$B\!=\!\gamma P^+_m\!+\!\delta (I_m\!-\!P^+_m)$
with $\alpha$, $\beta$, $\gamma$, and $\delta$
being real parameters.
The states invariant under $U\otimes U^*$-twirling are isotropic states,
but mixed states of full rank cannot be transformed to
$P^+_d$ as shown in the theorem 3 below.
The only isotropic state that is not full rank is $\varrho\!=\!P^+_m$,
and hence
$\gamma\!=\!0$ by $\hbox{tr}\varrho B\!=\!0$.
All the other conditions to be satisfied are
%%%%%%%%%%%%%%%%%%%%%%%%%%%%%%%%%%%%%%%%%%%%%%%%%%%%%%%%%%%%%%%%%%%%%%%%%%%%%%%
\begin{eqnarray}
1\ge\alpha>0, \hbox{~~}\beta\ge0, \hbox{~~}\delta\ge0,
\hbox{~~}1\ge \beta+\delta, && \cr
(d+1)\alpha+(d+1)(m-1)\beta+(m-1)\delta&\ge& 0, \cr
-(d+1)\alpha+(d+1)(m+1)\beta+(m+1)\delta&\ge& 0, \cr
-(d-1)\alpha-(d-1)(m-1)\beta+(m-1)\delta&\ge& 0, \cr
(d-1)\alpha-(d-1)(m+1)\beta+(m+1)\delta&\ge& 0,
\end{eqnarray}
%%%%%%%%%%%%%%%%%%%%%%%%%%%%%%%%%%%%%%%%%%%%%%%%%%%%%%%%%%%%%%%%%%%%%%%%%%%%%%%
for which solutions indeed exist for any $d$.
A solution which maximize $P(P^+_m\!\rightarrow\!P^+_d)\!=\!\alpha$ is
%%%%%%%%%%%%%%%%%%%%%%%%%%%%%%%%%%%%%%%%%%%%%%%%%%%%%%%%%%%%%%%%%%%%%%%%%%%%%%%
\begin{equation}
\left\{
\begin{array}{llll}
\alpha=\frac{m-1}{d-1}, & \beta=0, & \delta=1, &
\hbox{~for $d\!>\!m\!\ge\!2$,} \cr
\alpha=1, & \beta=0, & \delta=1, &
\hbox{~for $d\!\le\!m$.}
\end{array}
\right.
\end{equation}
%%%%%%%%%%%%%%%%%%%%%%%%%%%%%%%%%%%%%%%%%%%%%%%%%%%%%%%%%%%%%%%%%%%%%%%%%%%%%%%
Consequently, the following was obtained.
\par
%%%%%%%%%%%%%%%%%%%%%%%%%%%%%%%%%%%%%%%%%%%%%%%%%%%%%%%%%%%%%%%%%%%%%%%%%%%%%%%
%%
%%%%%%%%%%%%%%%%%%%%%%%%%%%%%%%%%%%%%%%%%%%%%%%%%%%%%%%%%%%%%%%%%%%%%%%%%%%%%%%
{Lemma 3:}
{\it
$P^+_m$ can be transformed to 
$P^+_d$ via SPPT maps with nonzero probability even when $d\!>\!m$.
The optimal probability is
$P(P^+_m\!\rightarrow\!P^+_d)\!=\!(m\!-\!1)/(d\!-\!1)$ for $d\!>\!m\!\ge\!2$
and $1$ for $d\!\le\!m$.
}
\par
%%%%%%%%%%%%%%%%%%%%%%%%%%%%%%%%%%%%%%%%%%%%%%%%%%%%%%%%%%%%%%%%%%%%%%%%%%%%%%%
%%
%%%%%%%%%%%%%%%%%%%%%%%%%%%%%%%%%%%%%%%%%%%%%%%%%%%%%%%%%%%%%%%%%%%%%%%%%%%%%%%
Then, let us consider the explicit method to implement the above
SPPT map.
It has been shown in \cite{Cirac01a} that
any PPT map can be implemented by SLOCC
assisted by a single copy of the state $E_{AB}$ in Eq.\ (\ref{eq: state E}).
The explicit form that implements $P^+_m\!\rightarrow\!P^+_d$ is
%%%%%%%%%%%%%%%%%%%%%%%%%%%%%%%%%%%%%%%%%%%%%%%%%%%%%%%%%%%%%%%%%%%%%%%%%%%%%%%
\begin{eqnarray}
E^{(1)}_{AB}&\!=\!&\frac{1}{m^2(d-1)}\big[
(m-1)
P^+_{dA_1B_1} \otimes P^+_{mA_2B_2} 
\label{eq: PPT-BE} \\
&&+
\frac{1}{d+1}
(I_d-P^+_d)_{A_1B_1} \otimes (I_m-P^+_m)_{A_2B_2} 
\big], \nonumber
\end{eqnarray}
%%%%%%%%%%%%%%%%%%%%%%%%%%%%%%%%%%%%%%%%%%%%%%%%%%%%%%%%%%%%%%%%%%%%%%%%%%%%%%%
which is not normalized since $S$ is not trace-preserving.
According to the implementation method of \cite{Cirac01a},
%%%%%%%%%%%%%%%%%%%%%%%%%%%%%%%%%%%%%%%%%%%%%%%%%%%%%%%%%%%%%%%%%%%%%%%%%%%%%%%
$$
\hbox{tr}_{23} P^+_{m A_2A_3} P^+_{m B_2B_3}
P^+_{mA_3B_3} E^{(1)}_{AB}=\frac{m-1}{m^4(d-1)}P^+_{dA_1B_1},
$$
%%%%%%%%%%%%%%%%%%%%%%%%%%%%%%%%%%%%%%%%%%%%%%%%%%%%%%%%%%%%%%%%%%%%%%%%%%%%%%%
and thus, the two Bell state measurements on
$\hbox{A}_2\hbox{A}_3$ and on $\hbox{B}_2\hbox{B}_3$ certainly yield
$P^+_{dA_1B_1}$,
when $P^+_{mA_3B_3}$ is an input state.
One may confirm that $E^{(1)}_{AB}$ is a PPT-state (and so
undistillable \cite{Horodecki98a}).
Generally, it is difficult to decide whether a given state is entangled or not,
but it is certain that $E^{(1)}_{AB}$ is entangled across the
$\hbox{A}_1\hbox{A}_2\!:\!\hbox{B}_1\hbox{B}_2$ cut,
since the transformation of $P^+_m\!\rightarrow\!P^+_d$,
which is impossible by SLOCC alone,
can be accomplished by the use of $E^{(1)}_{AB}$ as a resource, and hence:
\par
%%%%%%%%%%%%%%%%%%%%%%%%%%%%%%%%%%%%%%%%%%%%%%%%%%%%%%%%%%%%%%%%%%%%%%%%%%%%%%%
%%
%%%%%%%%%%%%%%%%%%%%%%%%%%%%%%%%%%%%%%%%%%%%%%%%%%%%%%%%%%%%%%%%%%%%%%%%%%%%%%%
{Lemma 4:}
{\it
$E^{(1)}_{AB}$ is a PPT-BE state for $d\!>\!m\!\ge\!2$. 
}
\par
%%%%%%%%%%%%%%%%%%%%%%%%%%%%%%%%%%%%%%%%%%%%%%%%%%%%%%%%%%%%%%%%%%%%%%%%%%%%%%%
%%
%%%%%%%%%%%%%%%%%%%%%%%%%%%%%%%%%%%%%%%%%%%%%%%%%%%%%%%%%%%%%%%%%%%%%%%%%%%%%%%
Now, the convertibility between arbitrary two bipartite pure states is clear.
Suppose that two distant parties initially share a pure entangled state
$|\psi_r\rangle$ with a Schmidt rank $r\!\ge\!2$,
and they wish to transform it to $|\phi_{r'}\rangle$
with a larger Schmidt rank $r'\!>\!r$
(the initial state must be entangled
since SPPT maps cannot transform separable states to entangled states).
First, they transform $|\psi_r\rangle$ to $P^+_r$ by appropriate SLOCC.
This is possible since the target state has the same Schmidt rank.
Then, they can transform it to $P^+_{r'}$ by SLOCC assisted by $E^{(1)}_{AB}$.
Finally, applying appropriate SLOCC to $P^+_{r'}$,
they can transform it to $|\phi_{r'}\rangle$.
On the other hand, the transformation decreasing the Schmidt rank is possible
by SLOCC alone, and hence the following was proved:
\par
%%%%%%%%%%%%%%%%%%%%%%%%%%%%%%%%%%%%%%%%%%%%%%%%%%%%%%%%%%%%%%%%%%%%%%%%%%%%%%%
%%
%%%%%%%%%%%%%%%%%%%%%%%%%%%%%%%%%%%%%%%%%%%%%%%%%%%%%%%%%%%%%%%%%%%%%%%%%%%%%%%
{Theorem 1:}
{\it
If two distant parties share appropriate PPT-BE states,
any bipartite pure entangled state can be transformed to any bipartite pure
state by SLOCC with nonzero probability.
}
\par
%%%%%%%%%%%%%%%%%%%%%%%%%%%%%%%%%%%%%%%%%%%%%%%%%%%%%%%%%%%%%%%%%%%%%%%%%%%%%%%
%%
%%%%%%%%%%%%%%%%%%%%%%%%%%%%%%%%%%%%%%%%%%%%%%%%%%%%%%%%%%%%%%%%%%%%%%%%%%%%%%%
It should be noted that, if $E^{(1)}_{AB}$ is normalized, the success
probability of the above implementation of $P^+_m\!\rightarrow\!P^+_d$ is
$1/[m^2(md\!+\!d\!-\!m)]$, which is much less than $(m\!-\!1)/(d\!-\!1)$
of the original SPPT map.
This is because only a single copy of $E^{(1)}_{AB}$
was used in the above implementation, while many (possibly infinitely many)
copies of PPT-BE states can be used as a resource in SPPT maps.
Although the implementation method using many copies of PPT-BE states
has not been known yet, the optimal probability may be worthwhile
for constructing a satisfying theory in the mathematical framework of PPT maps.
\par
%%%%%%%%%%%%%%%%%%%%%%%%%%%%%%%%%%%%%%%%%%%%%%%%%%%%%%%%%%%%%%%%%%%%%%%%%%%%%%%
%%
%%%%%%%%%%%%%%%%%%%%%%%%%%%%%%%%%%%%%%%%%%%%%%%%%%%%%%%%%%%%%%%%%%%%%%%%%%%%%%%
The optimal probability between two bipartite pure states by SLOCC
has been obtained in \cite{Vidal99a},
where it is determined by the ratio of the entanglement monotone that is
the partial summation of the squared Schmidt coefficients of pure states.
The above theorem shows that this quantity is not
monotone any more in PPT maps.
This suggests that even the majorization conditions
for the deterministic transformation of pure states \cite{Nielsen99a}
is not applicable to PPT maps (see the note of \cite{Note1}).
How the conditions are relaxed in PPT maps?
This seems to be an intriguing open problem.
It should be mentioned that
at least two convex functions, reduced von Neumann entropy and negativity
\cite{Vidal02a,Audenaert03a}, are monotone in deterministic and thus
trace-preserving PPT maps.
\par
%%%%%%%%%%%%%%%%%%%%%%%%%%%%%%%%%%%%%%%%%%%%%%%%%%%%%%%%%%%%%%%%%%%%%%%%%%%%%%%
%%
%%%%%%%%%%%%%%%%%%%%%%%%%%%%%%%%%%%%%%%%%%%%%%%%%%%%%%%%%%%%%%%%%%%%%%%%%%%%%%%
Let us next consider the stochastic transformation of tripartite states:
$|\psi_{ABC}\rangle\!\rightarrow\!|\phi_{ABC}\rangle$.
The dimension of each party (A, B or C) is arbitrary large but finite.
Suppose that $|\psi_{ABC}\rangle$ and $|\phi_{ABC}\rangle$ are genuine
tripartite entangled states.
Here, ``genuine'' means that those states have non-PPT with respect to
every bipartite partition (otherwise the pure state is biseparable or
fully separable).
Then, let us consider the unnormalized mixed state analogous to
$E^{(1)}_{AB}$ in Eq.\ (\ref{eq: PPT-BE}):
%%%%%%%%%%%%%%%%%%%%%%%%%%%%%%%%%%%%%%%%%%%%%%%%%%%%%%%%%%%%%%%%%%%%%%%%%%%%%%%
\begin{eqnarray}
E(x)_{ABC}&\!=\!&x
(|\phi\rangle\langle\phi|)_{A_1B_1C_1} 
\otimes 
(|\psi\rangle\langle\psi|)^T_{A_2B_2C_2} \\
&\!+\!&
(I\!-\!|\phi\rangle\langle\phi|)_{A_1B_1C_1}
\otimes
(I\!-\!|\psi\rangle\langle\psi|)^T_{A_2B_2C_2}, \nonumber
\end{eqnarray}
%%%%%%%%%%%%%%%%%%%%%%%%%%%%%%%%%%%%%%%%%%%%%%%%%%%%%%%%%%%%%%%%%%%%%%%%%%%%%%%
where $x$ is non-negative.
Let $|\psi^*\rangle\langle\psi^*|\!\equiv\!(|\psi\rangle\langle\psi|)^T$.
The Schmidt decomposition 
of $|\psi^*_{ABC}\rangle$ and $|\phi_{ABC}\rangle$
across the $\hbox{A}\!:\!\hbox{BC}$ cut is written as
$|\psi^*_{ABC}\rangle\!=\!\sum_{i}\sqrt{p_i}|i_{A}\rangle|i_{BC}\rangle$
and
$|\phi_{ABC}\rangle\!=\!\sum_{k}\sqrt{q_k}|k_{A}\rangle|k_{BC}\rangle$,
where $p_i$ and $q_k$ are chosen in decreasing order
such that $p_i\!\ge\!p_{i+1}$ and $q_k\!\ge\!q_{k+1}$.
Then, it is found that $[E(x)_{ABC}]^{T_A}\!\ge\!0$ if and only if
%%%%%%%%%%%%%%%%%%%%%%%%%%%%%%%%%%%%%%%%%%%%%%%%%%%%%%%%%%%%%%%%%%%%%%%%%%%%%%%
\begin{equation}
x\le \min \big[
\frac{1-p_1}{p_1}
\frac{1+\sqrt{q_1q_2}}{\sqrt{q_1q_2}},
\frac{1+\sqrt{p_1p_2}}{\sqrt{p_1p_2}}
\frac{1-q_1}{q_1}\big].
\label{eq: PPT condition for E}
\end{equation}
%%%%%%%%%%%%%%%%%%%%%%%%%%%%%%%%%%%%%%%%%%%%%%%%%%%%%%%%%%%%%%%%%%%%%%%%%%%%%%%
This is satisfied when $x\!=\!x_A\!\equiv\!(1\!-\!p_1)(1\!-\!q_1)/(p_1q_1)$,
which is nonzero positive
since $p_1\!<\!1$ and $q_1\!<\!1$ due to the assumption that
$(|\psi\rangle\langle\psi|)^{T_A}\!\not\ge\!0$
and
$(|\phi\rangle\langle\phi|)^{T_A}\!\not\ge\!0$.
Repeating this discussion for the other cuts $\hbox{B}\!:\!\hbox{AC}$ and
$\hbox{C}\!:\!\hbox{AB}$, obtaining $x_B$ and $x_C$, one can
choose $x_0\!=\!\min\{x_A,x_B,x_C\}\!>\!0$ such that
$E(x_0)_{ABC}$ has PPT with respect to every bipartite partition.
As a result, $E(x_0)_{ABC}$ is an separable or PPT-BE state,
since $E(x_0)_{ABC}$ is undistillable across every bipartite partition.
If three parties share $E(x_0)_{ABC}$,
they can accomplish the stochastic transformation of
$|\psi_{A_3B_3C_3}\rangle\!\rightarrow\!|\phi_{A_1B_1C_1}\rangle$
by multilocal operations similar to the bipartite case, i.e. three Bell
state measurements on $\hbox{A}_2\hbox{A}_3$, $\hbox{B}_2\hbox{B}_3$,
and $\hbox{C}_2\hbox{C}_3$.
The success probability of the transformation is nonzero since it is
proportional to $x_0/\hbox{tr}E(x_0)_{ABC}$.
In this way, if three parties share appropriate PPT-BE states,
they can transform a tripartite pure entangled state to arbitrary tripartite
pure entangled states at the single copy level.
This immediately implies that two genuine tripartite entangled states
of GHZ and W are inter-converted (Fig.\ \ref{fig: 2}).
This discussion can be extended to general multipartite settings in
a straightforward manner, and hence:
\par
%%%%%%%%%%%%%%%%%%%%%%%%%%%%%%%%%%%%%%%%%%%%%%%%%%%%%%%%%%%%%%%%%%%%%%%%%%%%%%%
%%
%%%%%%%%%%%%%%%%%%%%%%%%%%%%%%%%%%%%%%%%%%%%%%%%%%%%%%%%%%%%%%%%%%%%%%%%%%%%%%%
{Theorem 2:}
{\it
If $N$ distant parties share appropriate PPT-BE states,
all genuine $N$-partite pure entangled states are inter-convertible
by SLOCC at the single copy level.
}
\par
%%%%%%%%%%%%%%%%%%%%%%%%%%%%%%%%%%%%%%%%%%%%%%%%%%%%%%%%%%%%%%%%%%%%%%%%%%%%%%%
%%
%%%%%%%%%%%%%%%%%%%%%%%%%%%%%%%%%%%%%%%%%%%%%%%%%%%%%%%%%%%%%%%%%%%%%%%%%%%%%%%
This implies that the classification of $N$-partite entanglement is
drastically simplified when LOCC are assisted by PPT-bound entanglement:
all different types of $N$-partite entanglement are merged into only
one type.
\par
%%%%%%%%%%%%%%%%%%%%%%%%%%%%%%%%%%%%%%%%%%%%%%%%%%%%%%%%%%%%%%%%%%%%%%%%%%%%%%%
%%
%%%%%%%%%%%%%%%%%%%%%%%%%%%%%%%%%%%%%%%%%%%%%%%%%%%%%%%%%%%%%%%%%%%%%%%%%%%%%%%
One might expect that all non-PPT mixed states also
become inter-convertible at the single copy level when LOCC are assisted by
PPT-bound entanglement.
However, this is not the case.
Let us return to the bipartite settings and consider 
the transformation of $\varrho\!\rightarrow\!P^+_d$ when $\varrho$ is
a mixed state.
As shown in \cite{Kent98a},
SLOCC cannot transform a single copy of mixed states
$\varrho$ on ${\mathbb C}^m\otimes{\mathbb C}^m$
to pure entangled states
if $\hbox{rank}(\varrho)\!\ge\!m^2\!-\!2$.
PPT-BE states cannot improve the convertibility of such mixed states:
\par
%%%%%%%%%%%%%%%%%%%%%%%%%%%%%%%%%%%%%%%%%%%%%%%%%%%%%%%%%%%%%%%%%%%%%%%%%%%%%%%
%%
%%%%%%%%%%%%%%%%%%%%%%%%%%%%%%%%%%%%%%%%%%%%%%%%%%%%%%%%%%%%%%%%%%%%%%%%%%%%%%%
{Theorem 3:}
{\it
Even when two distant parties share PPT-BE states,
they cannot distill any pure entangled state from a single copy of
$\varrho$ on ${\mathbb C}^m\otimes{\mathbb C}^m$ if
$\hbox{rank}(\varrho)\!\ge\!m^2\!-\!2$.
}
\par
%%%%%%%%%%%%%%%%%%%%%%%%%%%%%%%%%%%%%%%%%%%%%%%%%%%%%%%%%%%%%%%%%%%%%%%%%%%%%%%
%%
%%%%%%%%%%%%%%%%%%%%%%%%%%%%%%%%%%%%%%%%%%%%%%%%%%%%%%%%%%%%%%%%%%%%%%%%%%%%%%%
Proof is rather involved but the key idea is as follows:
$\hbox{rank}(B)\!\le\!2$ when $\hbox{rank}(\varrho)\!\ge\!m^2\!-\!2$,
since $B\!\ge\!0$ and $\hbox{tr}\varrho B\!=\!0$.
On the other hand, $B^{T_A}\!\ge\!0$ must hold 
from $\frac{1}{d-1}B^{T_A}\!\ge\!-\frac{1}{d+1}B^{T_A}$,
and hence $B$ must be a separable state (leaving out normalization)
since $\hbox{rank}(B)\!\le\!m$ \cite{Horodecki00b}.
Therefore, by using appropriate local basis, $B$ must be written
as $B\!=\!y|11\rangle\langle11|+z|ef\rangle\langle ef|$ where
$|ef\rangle\!\equiv\!(\cos u|1\rangle\!+\!\sin u|2\rangle)\otimes
(\cos v|1\rangle\!+\!\sin v|2\rangle)$
is a product state.
Further, the support space of $A^{T_A}$ must be
contained in the support space of $B^{T_A}$ so that
$\frac{1}{d-1}B^{T_A}\!\ge\!A^{T_A}\!\ge\!-\frac{1}{d+1}B^{T_A}$.
Using the above form of $B$, one can see that 
$\hbox{tr}\varrho A\!>\!0$ is never satisfied, and
$\varrho\!\rightarrow\!P^+_d$ is
impossible for such $\varrho$ via any SPPT map.
This impossibility criterion immediately implies that
pure entangled states cannot be distilled from a single copy of
mixed states on ${\mathbb C}^2\otimes{\mathbb C}^2$
via any SPPT map.
\par
%%%%%%%%%%%%%%%%%%%%%%%%%%%%%%%%%%%%%%%%%%%%%%%%%%%%%%%%%%%%%%%%%%%%%%%%%%%%%%%
%%
%%%%%%%%%%%%%%%%%%%%%%%%%%%%%%%%%%%%%%%%%%%%%%%%%%%%%%%%%%%%%%%%%%%%%%%%%%%%%%%
In summary, I completely clarified the convertibility
between arbitrary multipartite pure states by
SLOCC with the assistance of PPT-BE states.
As a result, I showed that all $N$-partite pure entangled states are
inter-convertible by SLOCC at the single copy level, if $N$ distant parties
share appropriate PPT-BE states.
This implies that the Schmidt rank of a bipartite pure entangled state
can be increased, and that two incomparable tripartite entanglement of
the GHZ and W type can be inter-converted.
This is truly the effect of bound entanglement since the above entanglement
processing is impossible by SLOCC alone.
In this way, bound entanglement strongly influences the
convertibility of pure states.
However, there is still a limitation that bound entanglement cannot improve
the convertibility of some mixed states.
It will be important to further clarify the characteristics of hidden
resource, bound entanglement, to completely harness quantum power in
information processing.
\par
%%%%%%%%%%%%%%%%%%%%%%%%%%%%%%%%%%%%%%%%%%%%%%%%%%%%%%%%%%%%%%%%%%%%%%%%%%%%%%%
%%
%%%%%%%%%%%%%%%%%%%%%%%%%%%%%%%%%%%%%%%%%%%%%%%%%%%%%%%%%%%%%%%%%%%%%%%%%%%%%%%
The author would like to thank M.~B.~Plenio for helpful discussions.
%%%%%%%%%%%%%%%%%%%%%%%%%%%%%%%%%%%%%%%%%%%%%%%%%%%%%%%%%%%%%%%%%%%%%%%%%%%%%%%
%%
%%%%%%%%%%%%%%%%%%%%%%%%%%%%%%%%%%%%%%%%%%%%%%%%%%%%%%%%%%%%%%%%%%%%%%%%%%%%%%%
%\bibliography{personal}

\begin{thebibliography}{25}
\expandafter\ifx\csname natexlab\endcsname\relax\def\natexlab#1{#1}\fi
\expandafter\ifx\csname bibnamefont\endcsname\relax
  \def\bibnamefont#1{#1}\fi
\expandafter\ifx\csname bibfnamefont\endcsname\relax
  \def\bibfnamefont#1{#1}\fi
\expandafter\ifx\csname citenamefont\endcsname\relax
  \def\citenamefont#1{#1}\fi
\expandafter\ifx\csname url\endcsname\relax
  \def\url#1{\texttt{#1}}\fi
\expandafter\ifx\csname urlprefix\endcsname\relax\def\urlprefix{URL }\fi
\providecommand{\bibinfo}[2]{#2}
\providecommand{\eprint}[2][]{\url{#2}}

\bibitem[{\citenamefont{Bennett et~al.}(1996)\citenamefont{Bennett, Bernstein,
  Popescu, and Schumacher}}]{Bennett96b}
\bibinfo{author}{\bibfnamefont{C.~H.} \bibnamefont{Bennett}},
  \bibinfo{author}{\bibfnamefont{H.~J.} \bibnamefont{Bernstein}},
  \bibinfo{author}{\bibfnamefont{S.}~\bibnamefont{Popescu}}, \bibnamefont{and}
  \bibinfo{author}{\bibfnamefont{B.}~\bibnamefont{Schumacher}},
  \bibinfo{journal}{Phys. Rev. A} \textbf{\bibinfo{volume}{53}},
  \bibinfo{pages}{2046} (\bibinfo{year}{1996}).

\bibitem[{\citenamefont{Lo and Popescu}(2001)}]{Lo97a}
\bibinfo{author}{\bibfnamefont{H.~K.} \bibnamefont{Lo}} \bibnamefont{and}
  \bibinfo{author}{\bibfnamefont{S.}~\bibnamefont{Popescu}},
  \bibinfo{journal}{Phys. Rev. A} \textbf{\bibinfo{volume}{63}},
  \bibinfo{pages}{022301} (\bibinfo{year}{2001}).

\bibitem[{\citenamefont{Nielsen}(1999)}]{Nielsen99a}
\bibinfo{author}{\bibfnamefont{M.~A.} \bibnamefont{Nielsen}},
  \bibinfo{journal}{Phys. Rev. Lett.} \textbf{\bibinfo{volume}{83}},
  \bibinfo{pages}{436} (\bibinfo{year}{1999}).

\bibitem[{\citenamefont{Vidal}(1999)}]{Vidal99a}
\bibinfo{author}{\bibfnamefont{G.}~\bibnamefont{Vidal}},
  \bibinfo{journal}{Phys. Rev. Lett.} \textbf{\bibinfo{volume}{83}},
  \bibinfo{pages}{1046} (\bibinfo{year}{1999}).

\bibitem[{\citenamefont{D{\" u}r et~al.}(2000)\citenamefont{D{\" u}r, Vidal,
  and Cirac}}]{Dur00b}
\bibinfo{author}{\bibfnamefont{W.}~\bibnamefont{D{\" u}r}},
  \bibinfo{author}{\bibfnamefont{G.}~\bibnamefont{Vidal}}, \bibnamefont{and}
  \bibinfo{author}{\bibfnamefont{J.~I.} \bibnamefont{Cirac}},
  \bibinfo{journal}{Phys. Rev. A} \textbf{\bibinfo{volume}{62}},
  \bibinfo{pages}{062314} (\bibinfo{year}{2000}).

\bibitem[{\citenamefont{Horodecki et~al.}(1998)\citenamefont{Horodecki,
  Horodecki, and Horodecki}}]{Horodecki98a}
\bibinfo{author}{\bibfnamefont{M.}~\bibnamefont{Horodecki}},
  \bibinfo{author}{\bibfnamefont{P.}~\bibnamefont{Horodecki}},
  \bibnamefont{and}
  \bibinfo{author}{\bibfnamefont{R.}~\bibnamefont{Horodecki}},
  \bibinfo{journal}{Phys. Rev. Lett.} \textbf{\bibinfo{volume}{80}},
  \bibinfo{pages}{5239} (\bibinfo{year}{1998}).

\bibitem[{\citenamefont{Horodecki
  et~al.}(1999{\natexlab{a}})\citenamefont{Horodecki, Horodecki, and
  Horodecki}}]{Horodecki99a}
\bibinfo{author}{\bibfnamefont{M.}~\bibnamefont{Horodecki}},
  \bibinfo{author}{\bibfnamefont{P.}~\bibnamefont{Horodecki}},
  \bibnamefont{and}
  \bibinfo{author}{\bibfnamefont{R.}~\bibnamefont{Horodecki}},
  \bibinfo{journal}{Phys. Rev. A} \textbf{\bibinfo{volume}{60}},
  \bibinfo{pages}{1888} (\bibinfo{year}{1999}{\natexlab{a}}).

\bibitem[{\citenamefont{Horodecki et~al.}(2001)\citenamefont{Horodecki,
  Horodecki, Horodecki, Leung, and Terhal}}]{Horodecki01c}
\bibinfo{author}{\bibfnamefont{M.}~\bibnamefont{Horodecki}},
  \bibinfo{author}{\bibfnamefont{P.}~\bibnamefont{Horodecki}},
  \bibinfo{author}{\bibfnamefont{R.}~\bibnamefont{Horodecki}},
  \bibinfo{author}{\bibfnamefont{D.~W.} \bibnamefont{Leung}}, \bibnamefont{and}
  \bibinfo{author}{\bibfnamefont{B.~M.} \bibnamefont{Terhal}},
  \bibinfo{journal}{Quant. Inf. Comp.} \textbf{\bibinfo{volume}{1}},
  \bibinfo{pages}{70} (\bibinfo{year}{2001}).

\bibitem[{\citenamefont{Horodecki
  et~al.}(1999{\natexlab{b}})\citenamefont{Horodecki, Horodecki, and
  Horodecki}}]{Horodecki99c}
\bibinfo{author}{\bibfnamefont{P.}~\bibnamefont{Horodecki}},
  \bibinfo{author}{\bibfnamefont{M.}~\bibnamefont{Horodecki}},
  \bibnamefont{and}
  \bibinfo{author}{\bibfnamefont{R.}~\bibnamefont{Horodecki}},
  \bibinfo{journal}{Phys. Rev. Lett.} \textbf{\bibinfo{volume}{82}},
  \bibinfo{pages}{1056} (\bibinfo{year}{1999}{\natexlab{b}}).

\bibitem[{\citenamefont{Horodecki et~al.}()\citenamefont{Horodecki, Horodecki,
  Horodecki, and Oppenheim}}]{Horodecki03a}
\bibinfo{author}{\bibfnamefont{K.}~\bibnamefont{Horodecki}},
  \bibinfo{author}{\bibfnamefont{M.}~\bibnamefont{Horodecki}},
  \bibinfo{author}{\bibfnamefont{P.}~\bibnamefont{Horodecki}},
  \bibnamefont{and}
  \bibinfo{author}{\bibfnamefont{J.}~\bibnamefont{Oppenheim}},
  \bibinfo{journal}{quant-ph/0309110}.

\bibitem[{\citenamefont{Murao and Vedral}(2001)}]{Murao01a}
\bibinfo{author}{\bibfnamefont{M.}~\bibnamefont{Murao}} \bibnamefont{and}
  \bibinfo{author}{\bibfnamefont{V.}~\bibnamefont{Vedral}},
  \bibinfo{journal}{Phys. Rev. Lett.} \textbf{\bibinfo{volume}{86}},
  \bibinfo{pages}{352} (\bibinfo{year}{2001}).

\bibitem[{\citenamefont{D{\" u}r}(2001)}]{Dur01a}
\bibinfo{author}{\bibfnamefont{W.}~\bibnamefont{D{\" u}r}},
  \bibinfo{journal}{Phys. Rev. Lett.} \textbf{\bibinfo{volume}{87}},
  \bibinfo{pages}{230402} (\bibinfo{year}{2001}).

\bibitem[{\citenamefont{Shor et~al.}(2003)\citenamefont{Shor, Smolin, and
  Thapliyal}}]{Shor03a}
\bibinfo{author}{\bibfnamefont{P.~W.} \bibnamefont{Shor}},
  \bibinfo{author}{\bibfnamefont{J.~A.} \bibnamefont{Smolin}},
  \bibnamefont{and} \bibinfo{author}{\bibfnamefont{A.~V.}
  \bibnamefont{Thapliyal}}, \bibinfo{journal}{Phys. Rev. Lett.}
  \textbf{\bibinfo{volume}{90}}, \bibinfo{pages}{107901}
  (\bibinfo{year}{2003}).

\bibitem[{\citenamefont{D{\" u}r et~al.}(2004)\citenamefont{D{\" u}r, Cirac,
  and Horodecki}}]{Dur04a}
\bibinfo{author}{\bibfnamefont{W.}~\bibnamefont{D{\" u}r}},
  \bibinfo{author}{\bibfnamefont{J.~I.} \bibnamefont{Cirac}}, \bibnamefont{and}
  \bibinfo{author}{\bibfnamefont{P.}~\bibnamefont{Horodecki}},
  \bibinfo{journal}{Phys. Rev. Lett.} \textbf{\bibinfo{volume}{93}},
  \bibinfo{pages}{020503} (\bibinfo{year}{2004}).

\bibitem[{\citenamefont{Peres}(1996)}]{Peres96a}
\bibinfo{author}{\bibfnamefont{A.}~\bibnamefont{Peres}},
  \bibinfo{journal}{Phys. Rev. Lett.} \textbf{\bibinfo{volume}{77}},
  \bibinfo{pages}{1413} (\bibinfo{year}{1996}).

\bibitem[{\citenamefont{Rains}(1999)}]{Rains99b}
\bibinfo{author}{\bibfnamefont{E.~M.} \bibnamefont{Rains}},
  \bibinfo{journal}{Phys. Rev. A} \textbf{\bibinfo{volume}{60}},
  \bibinfo{pages}{173} (\bibinfo{year}{1999}).

\bibitem[{\citenamefont{Rains}(2001)}]{Rains01a}
\bibinfo{author}{\bibfnamefont{E.~M.} \bibnamefont{Rains}},
  \bibinfo{journal}{IEEE Trans. Inf. Theory} \textbf{\bibinfo{volume}{47}},
  \bibinfo{pages}{2921} (\bibinfo{year}{2001}).

\bibitem[{\citenamefont{Eggeling et~al.}(2001)\citenamefont{Eggeling,
  Vollbrecht, Werner, and Wolf}}]{Eggeling01a}
\bibinfo{author}{\bibfnamefont{T.}~\bibnamefont{Eggeling}},
  \bibinfo{author}{\bibfnamefont{K.~G.~H.} \bibnamefont{Vollbrecht}},
  \bibinfo{author}{\bibfnamefont{R.~F.} \bibnamefont{Werner}},
  \bibnamefont{and} \bibinfo{author}{\bibfnamefont{M.~M.} \bibnamefont{Wolf}},
  \bibinfo{journal}{Phys. Rev. Lett.} \textbf{\bibinfo{volume}{87}},
  \bibinfo{pages}{257902} (\bibinfo{year}{2001}).

\bibitem[{\citenamefont{Vollbrecht and Wolf}(2002)}]{Vollbrecht02a}
\bibinfo{author}{\bibfnamefont{K.~G.~H.} \bibnamefont{Vollbrecht}}
  \bibnamefont{and} \bibinfo{author}{\bibfnamefont{M.~M.} \bibnamefont{Wolf}},
  \bibinfo{journal}{Phys. Rev. Lett.} \textbf{\bibinfo{volume}{88}},
  \bibinfo{pages}{247901} (\bibinfo{year}{2002}).

\bibitem[{\citenamefont{Audenaert et~al.}(2003)\citenamefont{Audenaert, Plenio,
  and Eisert}}]{Audenaert03a}
\bibinfo{author}{\bibfnamefont{K.}~\bibnamefont{Audenaert}},
  \bibinfo{author}{\bibfnamefont{M.~B.} \bibnamefont{Plenio}},
  \bibnamefont{and} \bibinfo{author}{\bibfnamefont{J.}~\bibnamefont{Eisert}},
  \bibinfo{journal}{Phys. Rev. Lett.} \textbf{\bibinfo{volume}{90}},
  \bibinfo{pages}{027901} (\bibinfo{year}{2003}).

\bibitem[{\citenamefont{Cirac et~al.}(2001)\citenamefont{Cirac, D{\" u}r,
  Kraus, and Lewenstein}}]{Cirac01a}
\bibinfo{author}{\bibfnamefont{J.~I.} \bibnamefont{Cirac}},
  \bibinfo{author}{\bibfnamefont{W.}~\bibnamefont{D{\" u}r}},
  \bibinfo{author}{\bibfnamefont{B.}~\bibnamefont{Kraus}}, \bibnamefont{and}
  \bibinfo{author}{\bibfnamefont{M.}~\bibnamefont{Lewenstein}},
  \bibinfo{journal}{Phys. Rev. Lett.} \textbf{\bibinfo{volume}{86}},
  \bibinfo{pages}{544} (\bibinfo{year}{2001}).

\bibitem{Note1}
M.~B.~Plenio and J.~Eisert have found explicit examples where $P^+_m$ is
deterministically transformed to
some pure entangled states with higher Schmidt rank (and with
smaller negativity) using trace-preserving PPT maps shown in
\protect\cite{Audenaert03a} (private communication).

\bibitem[{\citenamefont{Vidal and Werner}(2002)}]{Vidal02a}
\bibinfo{author}{\bibfnamefont{G.}~\bibnamefont{Vidal}} \bibnamefont{and}
  \bibinfo{author}{\bibfnamefont{R.~F.} \bibnamefont{Werner}},
  \bibinfo{journal}{Phys. Rev. A} \textbf{\bibinfo{volume}{65}},
  \bibinfo{pages}{032314} (\bibinfo{year}{2002}).

\bibitem[{\citenamefont{Kent}(1998)}]{Kent98a}
\bibinfo{author}{\bibfnamefont{A.}~\bibnamefont{Kent}}, \bibinfo{journal}{Phys.
  Rev. Lett.} \textbf{\bibinfo{volume}{81}}, \bibinfo{pages}{2839}
  (\bibinfo{year}{1998}).

\bibitem[{\citenamefont{Horodecki et~al.}(2000)\citenamefont{Horodecki,
  Lewenstein, Vidal, and Cirac}}]{Horodecki00b}
\bibinfo{author}{\bibfnamefont{P.}~\bibnamefont{Horodecki}},
  \bibinfo{author}{\bibfnamefont{M.}~\bibnamefont{Lewenstein}},
  \bibinfo{author}{\bibfnamefont{G.}~\bibnamefont{Vidal}}, \bibnamefont{and}
  \bibinfo{author}{\bibfnamefont{I.}~\bibnamefont{Cirac}},
  \bibinfo{journal}{Phys. Rev. A} \textbf{\bibinfo{volume}{62}},
  \bibinfo{pages}{032310} (\bibinfo{year}{2000}).

\end{thebibliography}
%\bibliographystyle{apsrev}

%%%%%%%%%%%%%%%%%%%%%%%%%%%%%%%%%%%%%%%%%%%%%%%%%%%%%%%%%%%%%%%%%%%%%%%%%%%%%%%
%%
%%%%%%%%%%%%%%%%%%%%%%%%%%%%%%%%%%%%%%%%%%%%%%%%%%%%%%%%%%%%%%%%%%%%%%%%%%%%%%%
%
%
%\end{document}
%
% The followings are added for referring purpose.
%
%%%%%%%%%%%%%%%%%%%%%%%%%%%%%%%%%%%%%%%%%%%%%%%%%%%%%%%%%%%%%%%%%%%%%%%%%%%%%%%
%%
%%%%%%%%%%%%%%%%%%%%%%%%%%%%%%%%%%%%%%%%%%%%%%%%%%%%%%%%%%%%%%%%%%%%%%%%%%%%%%%
\section{APPENDIX A: Proof of the theorem 3}
In this appendix, the complete proof of the theorem 3 is given.
Since $B\!\ge\!0$ and $\hbox{tr}\varrho B\!=\!0$, the support space of $B$
must be contained in the kernel space of $\varrho$, and hence
$\hbox{rank}(B)\!\le\!2$
when $\hbox{rank}(\varrho)\!\ge\!m^2\!-\!2$.
On the other hand, $B^{T_A}\!\ge\!0$ must hold 
from $\frac{1}{d-1}B^{T_A}\!\ge\!-\frac{1}{d+1}B^{T_A}$,
and $B$ must be a separable state (leaving out normalization)
since $\hbox{rank}(B)\!\le\!m$ \cite{Horodecki00b}.
Therefore, by using appropriate local basis, $B$ can be written
as
%%%%%%%%%%%%%%%%%%%%%%%%%%%%%%%%%%%%%%%%%%%%%%%%%%%%%%%%%%%%%%%%%%%%%%%%%%%%%%%
\begin{equation}
B=y|11\rangle\langle11|+z|ef\rangle\langle ef|
\end{equation}
%%%%%%%%%%%%%%%%%%%%%%%%%%%%%%%%%%%%%%%%%%%%%%%%%%%%%%%%%%%%%%%%%%%%%%%%%%%%%%%
where $y$ and $z$ are non-negative values and
%%%%%%%%%%%%%%%%%%%%%%%%%%%%%%%%%%%%%%%%%%%%%%%%%%%%%%%%%%%%%%%%%%%%%%%%%%%%%%%
\begin{equation}
|ef\rangle=(\cos u|1\rangle+\sin u|2\rangle)\otimes
(\cos v|1\rangle+\sin v|2\rangle)
\end{equation}
%%%%%%%%%%%%%%%%%%%%%%%%%%%%%%%%%%%%%%%%%%%%%%%%%%%%%%%%%%%%%%%%%%%%%%%%%%%%%%%
is a product vector.
In this choice of local basis, $B^{T_A}\!=\!B$.
Let $P$ be the projector on the support space of $B^{T_A}$ and
$Q\!\equiv\!I\!-\!P$.
The condition of
$\frac{1}{d-1}B^{T_A}\!\ge\!A^{T_A}\!\ge\!-\frac{1}{d+1}B^{T_A}$
implies that $\pm QA^{T_A}Q\!\ge\!0$, and hence $QA^{T_A}Q\!=\!0$ must hold.
Further,
$A^{T_A}+\frac{1}{d+1}B^{T_A}$ must be a positive operator,
for which $Q(A^{T_A}\!+\!\frac{1}{d+1}B^{T_A})Q\!=\!0$ also holds.
Therefore, support space of $A^{T_A}\!+\!\frac{1}{d+1}B^{T_A}$ must be
$P$, and hence the support space of $A^{T_A}$ must be contained in the
support space of $B^{T_A}$.
As a result, the rank of $A^{T_A}$ is at most 2.
Further, $A^{T_A}$ must be written in the form of
%%%%%%%%%%%%%%%%%%%%%%%%%%%%%%%%%%%%%%%%%%%%%%%%%%%%%%%%%%%%%%%%%%%%%%%%%%%%%%%
\begin{equation}
A^{T_A}=r|11\rangle\langle 11|+s|11\rangle\langle ef|+s^* |ef\rangle\langle 11|
+t|ef\rangle\langle ef|,
\end{equation}
%%%%%%%%%%%%%%%%%%%%%%%%%%%%%%%%%%%%%%%%%%%%%%%%%%%%%%%%%%%%%%%%%%%%%%%%%%%%%%%
and $A$ is given by
%%%%%%%%%%%%%%%%%%%%%%%%%%%%%%%%%%%%%%%%%%%%%%%%%%%%%%%%%%%%%%%%%%%%%%%%%%%%%%%
\begin{equation}
A=r|11\rangle\langle 11|+s|e1\rangle\langle 1f|+s^* |1f\rangle\langle e1|
+t|ef\rangle\langle ef|.
\end{equation}
%%%%%%%%%%%%%%%%%%%%%%%%%%%%%%%%%%%%%%%%%%%%%%%%%%%%%%%%%%%%%%%%%%%%%%%%%%%%%%%
Therefore, $A$ must be essentially two-qubit state (leaving out normalization)
since $A\!\ge\!0$ must hold according to the lemma 1.
If the two-qubit state $A$ is entangled,
$A^{T_A}$ must be rank 4 
\footnote{S.~Ishizaka, Phys. Rev. A {\bf 69}, 020301 (2004)},
which contradicts that the rank of $A^{T_A}$ is at most 2.
Therefore, $A$ and $A^{T_A}$ must be written in a separable form.
\par
%%%%%%%%%%%%%%%%%%%%%%%%%%%%%%%%%%%%%%%%%%%%%%%%%%%%%%%%%%%%%%%%%%%%%%%%%%%%%%%
%
%%%%%%%%%%%%%%%%%%%%%%%%%%%%%%%%%%%%%%%%%%%%%%%%%%%%%%%%%%%%%%%%%%%%%%%%%%%%%%%
In the case where $\sin u \sin v\!\ne\!0$, the support space of $A^{T_A}$
spanned by $|11\rangle$ and $|ef\rangle$ contains only two product vectors
($|11\rangle$ and $|ef\rangle$ itself)
\footnote{A.~Sanpera, {\it et. al.}, Phys. Rev. A {\bf 58}, 826 (1998)},
and hence $A^{T_A}$ must be written as
%%%%%%%%%%%%%%%%%%%%%%%%%%%%%%%%%%%%%%%%%%%%%%%%%%%%%%%%%%%%%%%%%%%%%%%%%%%%%%%
\begin{equation}
A^{T_A}=r|11\rangle\langle 11|+t|ef\rangle\langle ef|=A.
\end{equation}
%%%%%%%%%%%%%%%%%%%%%%%%%%%%%%%%%%%%%%%%%%%%%%%%%%%%%%%%%%%%%%%%%%%%%%%%%%%%%%%
As a result, the support space of $A$ is contained in the support space
of $B$ and $\hbox{tr}\varrho A\!>\!0$ is never satisfied.
In the case where $\sin u \sin v\!=\!0$, $|e\rangle\!=\!|1\rangle$
or $|f\rangle\!=\!|1\rangle$ holds.
As a result, $A$ is spanned by $\{|11\rangle,|1f\rangle\}$
(or $\{|11\rangle,|e1\rangle\}$) and
$\hbox{tr}\varrho A\!>\!0$ is never satisfied again,
since $\{|11\rangle,|1f\rangle\}$ (or $\{|11\rangle,|e1\rangle\}$) are kernels
of $\varrho$.
%%%%%%%%%%%%%%%%%%%%%%%%%%%%%%%%%%%%%%%%%%%%%%%%%%%%%%%%%%%%%%%%%%%%%%%%%%%%%%%
%%
%%%%%%%%%%%%%%%%%%%%%%%%%%%%%%%%%%%%%%%%%%%%%%%%%%%%%%%%%%%%%%%%%%%%%%%%%%%%%%%
%%%%%%%%%%%%%%%%%%%%%%%%%%%%%%%%%%%%%%%%%%%%%%%%%%%%%%%%%%%%%%%%%%%%%%%%%%%%%%%
%%
%%%%%%%%%%%%%%%%%%%%%%%%%%%%%%%%%%%%%%%%%%%%%%%%%%%%%%%%%%%%%%%%%%%%%%%%%%%%%%%
\section{APPENDIX B: Positivity of the partial transpose of $E(x)_{ABC}$}
\label{sec: Positivity of the partial transpose of E}
In this appendix, the necessary and sufficient condition
Eq.\ (\ref{eq: PPT condition for E}) for $[E_{ABC}(x)]^{T_A}\!\ge\!0$
is derived.
All indices denoting parties are omitted for simplicity as
%%%%%%%%%%%%%%%%%%%%%%%%%%%%%%%%%%%%%%%%%%%%%%%%%%%%%%%%%%%%%%%%%%%%%%%%%%%%%%%
\begin{eqnarray}
E(x)_{ABC}&\!=\!&x
|\phi\rangle\langle\phi|\otimes |\psi^*\rangle\langle\psi^*| \cr
&\!+\!&
\big(I-|\phi\rangle\langle\phi|\big)
\otimes
\big(I-|\psi^*\rangle\langle\psi^*|\big).
\end{eqnarray}
%%%%%%%%%%%%%%%%%%%%%%%%%%%%%%%%%%%%%%%%%%%%%%%%%%%%%%%%%%%%%%%%%%%%%%%%%%%%%%%
Similarly, $|\psi^*\rangle$ and $|\phi\rangle$ are written as
%%%%%%%%%%%%%%%%%%%%%%%%%%%%%%%%%%%%%%%%%%%%%%%%%%%%%%%%%%%%%%%%%%%%%%%%%%%%%%%
\begin{eqnarray}
|\psi^*\rangle&\!=\!&\sum_i\sqrt{p_i}|ii\rangle,\cr
|\phi\rangle&\!=\!&\sum_k\sqrt{q_k}|kk\rangle.
\end{eqnarray}
%%%%%%%%%%%%%%%%%%%%%%%%%%%%%%%%%%%%%%%%%%%%%%%%%%%%%%%%%%%%%%%%%%%%%%%%%%%%%%%
The partial transpose of $|\psi^*\rangle$ and $|\phi\rangle$ is given by
%%%%%%%%%%%%%%%%%%%%%%%%%%%%%%%%%%%%%%%%%%%%%%%%%%%%%%%%%%%%%%%%%%%%%%%%%%%%%%%
\begin{eqnarray}
(|\psi^*\rangle\langle\psi^*|)^{T_A}&=&
\sum_i p_i |ii\rangle\langle ii|
+\sum_{j>i}\sqrt{p_ip_j}|\psi^+_{ij}\rangle\langle\psi^+_{ij}|\cr
&&-\sum_{j>i}\sqrt{p_ip_j}|\psi^-_{ij}\rangle\langle\psi^-_{ij}|, \cr
(|\phi\rangle\langle\phi|)^{T_A}&=&
\sum_k q_k |kk\rangle\langle kk|
+\sum_{l>k}\sqrt{q_kq_l}|\psi^+_{kl}\rangle\langle\psi^+_{kl}| \cr
&&-\sum_{l>k}\sqrt{q_kq_l}|\psi^-_{kl}\rangle\langle\psi^-_{kl}|,
\end{eqnarray}
%%%%%%%%%%%%%%%%%%%%%%%%%%%%%%%%%%%%%%%%%%%%%%%%%%%%%%%%%%%%%%%%%%%%%%%%%%%%%%%
where $|\psi^{\pm}_{ij}\rangle=(|ij\rangle\pm|ij\rangle)/\sqrt{2}$.
The partial transpose of $E(x)_{ABC}$ is 
%%%%%%%%%%%%%%%%%%%%%%%%%%%%%%%%%%%%%%%%%%%%%%%%%%%%%%%%%%%%%%%%%%%%%%%%%%%%%%%
\begin{eqnarray}
[E(x)_{ABC}]^{T_A}
&=&\sum_k |kk\rangle\langle kk| \otimes 
\Big[x q_k(|\psi^*\rangle\langle\psi^*|)^{T_A} \cr
&&+(1-q_k)(I-|\psi^*\rangle\langle\psi^*|)^{T_A}\Big] \cr
&+&\sum_{l>k} |\psi^+_{kl}\rangle\langle\psi^+_{kl}| \otimes 
 \Big[x \sqrt{q_kq_l}(|\psi^*\rangle\langle\psi^*|)^{T_A} \cr
&&+(1-\sqrt{q_kq_l})(I-|\psi^*\rangle\langle\psi^*|)^{T_A}\Big] \cr
&+&\sum_{l>k}|\psi^-_{kl}\rangle\langle\psi^-_{kl}|\otimes 
\Big[-x \sqrt{q_kq_l}(|\psi^*\rangle\langle\psi^*|)^{T_A} \cr
&&+(1+\sqrt{q_kq_l})(I-|\psi^*\rangle\langle\psi^*|)^{T_A}\Big].
\end{eqnarray}
%%%%%%%%%%%%%%%%%%%%%%%%%%%%%%%%%%%%%%%%%%%%%%%%%%%%%%%%%%%%%%%%%%%%%%%%%%%%%%%
Therefore, $[E(x)_{ABC}]^{T_A}\!\ge\!0$ if and only if
%%%%%%%%%%%%%%%%%%%%%%%%%%%%%%%%%%%%%%%%%%%%%%%%%%%%%%%%%%%%%%%%%%%%%%%%%%%%%%%
\begin{eqnarray}
\big[(x\!+\!1)q_k\!-\!1\big](|\psi^*\rangle\langle\psi^*|)^{T_A}
\!+\!(1\!-\!q_k)I&\ge&0, \cr
\big[(x\!+\!1)\sqrt{q_kq_l}\!-\!1\big](|\psi^*\rangle\langle\psi^*|)^{T_A}
\!+\!(1\!-\!\sqrt{q_kq_l})I&\ge& 0, \cr
-\big[(x\!+\!1)\sqrt{q_kq_l}\!+\!1\big](|\psi^*\rangle\langle\psi^*|)^{T_A}
\!+\!(1\!+\!\sqrt{q_iq_j})I&\ge& 0, \cr
&&
\end{eqnarray}
%%%%%%%%%%%%%%%%%%%%%%%%%%%%%%%%%%%%%%%%%%%%%%%%%%%%%%%%%%%%%%%%%%%%%%%%%%%%%%%
for all $l\!>\!k$.
Since $x$ is a non-negative real parameter, those conditions are
satisfied if and only if
%%%%%%%%%%%%%%%%%%%%%%%%%%%%%%%%%%%%%%%%%%%%%%%%%%%%%%%%%%%%%%%%%%%%%%%%%%%%%%%
\begin{eqnarray}
-\big[(x+1) q_k -1\big]\sqrt{p_iq_j}+1-q_k&\ge& 0, \cr
-\big[(x+1) \sqrt{q_kq_l} -1\big]\sqrt{p_ip_j}+1-\sqrt{q_kq_l}&\ge& 0, \cr
-\big[(x+1)\sqrt{q_kq_l}+1\big]p_i+1+\sqrt{q_kq_l}&\ge& 0, \cr
-\big[(x+1)\sqrt{q_kq_l}+1\big]\sqrt{p_ip_j}+1+\sqrt{q_kq_l}&\ge& 0,
\end{eqnarray}
%%%%%%%%%%%%%%%%%%%%%%%%%%%%%%%%%%%%%%%%%%%%%%%%%%%%%%%%%%%%%%%%%%%%%%%%%%%%%%%
for all $j\!>\!i$ and $k\!>\!l$, which lead to
%%%%%%%%%%%%%%%%%%%%%%%%%%%%%%%%%%%%%%%%%%%%%%%%%%%%%%%%%%%%%%%%%%%%%%%%%%%%%%%
\begin{eqnarray}
x&\le& \min_{j>i,k>l} \bigg[
\frac{1-\sqrt{p_ip_j}}{\sqrt{p_ip_j}}
\frac{1+\sqrt{q_kq_l}}{\sqrt{q_kq_l}},
\frac{1+\sqrt{p_ip_j}}{\sqrt{p_ip_j}}
\frac{1-\sqrt{q_kq_l}}{\sqrt{q_kq_l}}, \cr
&&\hbox{~~~~~~~}\frac{1-p_i}{p_i}
\frac{1+\sqrt{q_kq_l}}{\sqrt{q_kq_l}},
\frac{1+\sqrt{p_ip_j}}{\sqrt{p_ip_j}}
\frac{1-q_k}{q_k}\bigg].
\end{eqnarray}
%%%%%%%%%%%%%%%%%%%%%%%%%%%%%%%%%%%%%%%%%%%%%%%%%%%%%%%%%%%%%%%%%%%%%%%%%%%%%%%
Since $p_1$ and $q_1$ are largest among $p_i$ and $q_i$, respectively,
the above condition is satisfied if and only if
Eq.\ (\ref{eq: PPT condition for E}) is satisfied.
\par
%%%%%%%%%%%%%%%%%%%%%%%%%%%%%%%%%%%%%%%%%%%%%%%%%%%%%%%%%%%%%%%%%%%%%%%%%%%%%%%
%%
%%%%%%%%%%%%%%%%%%%%%%%%%%%%%%%%%%%%%%%%%%%%%%%%%%%%%%%%%%%%%%%%%%%%%%%%%%%%%%%

\end{document}